\documentclass[%
 reprint, 
 amsmath,amssymb,
 aps,
 nofootinbib
 prb,
 superscriptaddress
 ]{revtex4-1}
\usepackage{scrextend}
\usepackage{graphicx}
\usepackage{dcolumn}
\usepackage{bm}
\usepackage{tabularx}
\usepackage[utf8]{inputenc}
\usepackage[english]{babel}
\usepackage[bottom]{footmisc}
\usepackage{color}

\usepackage{amsmath}
\usepackage{soul,xcolor}
\setstcolor{red}
\usepackage{hyperref}
\usepackage{booktabs}

\DeclareMathAlphabet{\pazocal}{OMS}{zplm}{m}{n}

\begin{document}

\preprint{APS/123-QED}

\title{\textit{Ab initio} study of NaSrSb and NaBaSb as potential thermoelectric
prospects}

\author{Chandan Kumar Vishwakarma}
\affiliation{Department of Physics, Indian Institute of Technology,
             Hauz Khas, New Delhi 110016, India}
\author{Mohd Zeeshan}
\affiliation{Department of Physics, Indian Institute of Technology,
             Hauz Khas, New Delhi 110016, India}
\author{B. K. Mani}
\email{bkmani@physics.iitd.ac.in}
\affiliation{Department of Physics, Indian Institute of Technology,
Hauz Khas, New Delhi 110016, India}

\date{\today}
             
\begin{abstract} 
Zintl phases are excellent thermoelectric prospects to put
the waste heat to good use. In the quest of the same, using
first-principles methods combined with Boltzmann transport
theory, we explored two recent phases NaSrSb and NaBaSb. We
found low lattice thermal conductivity of 1.9
and 1.3 W m$^{-1}$ K$^{-1}$ at 300~K for NaSrSb and NaBaSb,
respectively, which are of the same order as other potential Zintl
phases such as Sr$_3$AlSb$_3$ and BaCuSb. We account for such low
values to short phonon lifetimes, small phonon group velocities,
and lattice anharmonicity in the crystal structure. The calculated
electrical transport parameters based on acoustic deformation potential,
ionized impurity, and polar optical phonon scattering mechanisms reveal
large Seebeck coefficients for both materials. Further, we obtain a high
figure of merit of ZT$\sim$2.0 at 900~K for \textit{n}-type NaSrSb. On
the other hand, the figure of merit of \textit{n}-type NaBaSb surpasses
the unity. We are optimistic about our findings and believe our work
would set a basis for future experimental investigations.
\end{abstract}

\maketitle

\section{Introduction}
The search for new materials has always fascinated 
researchers around the globe, as the new materials
can be explored in a various technological applications.
Advances in synthesis techniques and experimentalists'
persistent endeavors have led to the discovery of
numerous materials. Though, sometimes the motive of
the discovery is to document the missing materials.
However, it is also not always possible to thoroughly
explore the material owing to the challenges in
characterization and difficulties of sample handling.
In such cases, first-principles simulations can play
a vital role in addressing the invaluable missing
information. The prediction of materials properties
utilizing first-principles simulations could greatly
facilitate understanding of the nature of material
and may attract experimentalists' attention.

Recently, Wang and Bobev synthesized NaSrSb, NaBaSb,
and NaEuSb, and their preliminary calculations hinted
at features of topological quantum materials
\cite{Wang23}.
Further, the density of states and crystal orbital Hamilton
curves were suggestive of good thermoelectric properties.
Importantly, it was suggested that the materials belong to classical
Zintl phases, which are well-acknowledged representatives of thermoelectric
materials \cite{Guo23}. Zintl phases are known for their complex structures,
hierarchical bonding, low sound velocities, and lattice anharmonicity,
which accounts for low lattice thermal conductivity -- a key ingredient
for a good thermoelectric material \cite{Jana16, Jana17, Kim00, Huang18}. In addition, these phases can
offer good electronic properties arising from dispersive bands \cite{Gorai20}.
Thus, it will be interesting to investigate the aforementioned
materials for thermoelectric applications.

To our knowledge, a literature review reveals that
NaBaSb and NaEuSb still need to be explored for thermoelectrics. On the other
hand, in the case of NaSrSb, the first-principles calculations for thermoelectric
properties are reported \cite{Chepkoech22}. Though the authors have well-explored the thermal
properties, the electronic transport properties are calculated utilizing
relaxation time derived through Bardeen and Shockley approach \cite{Bardeen50}. The formalism
has been successfully used in many systems \cite{Rugut19, Mahmoud19}. However, the deformation potential
scattering due to long wavelength acoustic phonons is treated using only an
averaged elastic constant and band edge deformation potential. The downside
of the method is that the crucial perturbations from transverse phonon modes
and anisotropy in the deformation response are not considered. Since
relaxation time significantly affects the charge carriers' mobility, computing accurate carrier lifetimes using more sophisticated
formalism could help extract reliable transport properties.
Recently, a more computationally efficient method for calculating the scattering rates has been devised, extending the existing methods for isotropic band structures to anisotropic materials \cite{Ganose21}. Because NaSrSb belongs to the anisotropic category,
more reliable electrical transport properties can be calculated, which could be
beneficial in accurately governing the thermoelectric potential of NaSrSb.

%

In this paper, we have demonstrated the thermoelectric potential
of NaSrSb and NaBaSb utilizing first-principles simulations.
We found remarkably low lattice thermal conductivity for both
materials. The origin of such low values is traced to
short phonon lifetimes, low group velocities, and lattice
anharmonicity. In addition, we have studied the electronic structure
and calculated the scattering rates and mobility of charge carriers
to extract the electrical transport coefficients. We obtained a reasonably
good power factor for NaSrSb at optimal carrier concentration.
Based on low lattice thermal conductivity and high power factor,
we found an excellent figure of merit for NaSrSb which is in-line
with some other state-of-the-art thermoelectric materials. Further,
we obtained a figure of merit exceeding unity for NaBaSb.

The paper is organized as follows: Sec. II discusses the
computational methods used, Sec. III presents the results of
thermal transport, electronic structure, and electrical transport.
Finally, we summarize the paper in Sec. IV.

\section{Computational Methods}

Our first-principles calculations are performed
within the framework of density functional theory
as implemented in the Vienna Ab Initio Simulation
Package (VASP), using projector-augmented wave
pseudopotentials \cite{Kresse96, KressePRB}. The generalized gradient
approximation (GGA) with the Perdew-Burke-Ernzerhof
(PBE) exchange-correlation functional is adopted \cite{Perdew96}.
The cutoff energy for the plane waves is set as 500 eV.
The crystal structure is fully optimized by sampling
the Brillouin zone with a Monkhorst-Pack \textit{k} mesh
of 11$\times$11$\times$11 whereas a denser mesh of
21$\times$21$\times$21 is used for self-consistent-field
electronic calculations. The convergence criteria for
the total energy and atomic forces are 10$^{-8}$ eV and
10$^{-7}$ eV/Å, respectively. In order to acquire a reliable
band gap, we further employed the modified Becke-Johnson (mBJ)
potential. Since the chosen materials comprise heavy atoms, we also performed the full-relativistic calculations
to see the effect of spin-orbit coupling (SOC). 

We constructed a 2$\times$2$\times$2 supercell (72 atoms)
with default finite atomic displacements to collect the second-order
harmonic and third-order anharmonic interatomic force constants.
The second-order force constants and phonons are computed using the
Phonopy package \cite{Togo15}. The phonon dispersion curves are obtained by solving
the equation
\begin{equation}
  \sum_{\beta\tau'} D^{\alpha\beta}_{\tau \tau'}
	(\mathbf{q}) \gamma^{\beta\tau'}_{\mathbf{q}j} = 
	\omega^2_{\mathbf{q}j}\gamma^{\alpha\tau}_{\mathbf{q}j}. 
\end{equation}
where the indices $\tau, \tau'$ represent the atoms, 
$\alpha, \beta$ are the Cartesian coordinates, ${\mathbf{q}}$ is
a wave vector, $j$ is a band index, $D(\mathbf{q})$ represents
the dynamical matrix, $\omega$ signifies the corresponding 
phonon frequency, and $\gamma$ is the polarization vector.
Mode Gr{\"u}neisen parameter is calculated by performing
phonon calculations at three different volumes, viz. 
equilibrium unit-cell volume, slightly larger (2\%) and smaller volume
(2\%), whereas mean square displacements of the atoms are derived
from the number of phonon excitations. 

For the evaluation of third-order force constants and lattice thermal
conductivity, we have used the Phono3py package \cite{TogoPRB}. The lattice thermal
conductivity is obtained by using the direct solution of linearized
phonon Boltzmann equation and is expressed as
\begin{equation}
   \kappa_L = \frac{1}{NV} \sum_\lambda C_\lambda \mathbf {v}_\lambda 
              \otimes \mathbf {v}_\lambda \tau_\lambda
\end{equation}
where $N$, $V$, $C$, $\mathbf {v}$, $\tau$, and the index $\lambda$
represent the number of unit-cells in the supercell, volume of the unit-cell, heat capacity, phonon group velocity, single-mode relaxation time, and the phonon mode, respectively. 
We have not included any crude force-constant
approximation, however, we have tested the convergence of lattice thermal
conductivity in terms of increasing cutoff distance. We found that a
cutoff distance of 7 Å yields a negligible change in the values. Albeit,
it required performing calculations for 1326 displacements for each system.
We further assessed the lattice thermal conductivity convergence with respect to \textit{q} grid and found that a grid of 21$\times$21$\times$21
is grossly sufficient. 

The electrical transport calculations are performed using the formalism
implemented in the Amset code \cite{Ganose21}. The components of scattering time
for electrons from an initial state $n$\textbf{k} to final state
$m$\textbf{k} + $q$ are calculated using Fermi’s golden rule
\begin{equation}
    \tilde{\tau}^{-1}_{n \mathbf{k} \rightarrow m \mathbf{k} + q} = 
    \frac{2 \pi}{\hbar} \arrowvert g_{nm} (\mathbf{k},\mathbf{q}) \arrowvert^2 \delta (\epsilon_{n \mathbf{k}} - \epsilon_{m \mathbf{k} + q}) 
\end{equation}
where $\hbar$ is the reduced Planck’s constant, $\epsilon$ is the electron energy, $\delta$ is the Dirac delta function and $g$ is the coupling matrix element. The different scattering mechanisms considered for the 
matrix element are acoustic deformation potential (ADP), ionized
impurity (IMP), and polar-optical phonon (POP) mechanism. The required
parameters for different scattering mechanisms such as elastic and
dielectric constants, deformation potential, and polar optical phonon
frequency, are extracted from first-principles calculations. The Seebeck coefficient, electrical conductivity, and electronic thermal conductivity are obtained by solving the Boltzmann transport equation
implemented in the Amset code. The transport properties are well
tested for convergence in terms of interpolation factor. A default
interpolation factor of 10 provides satisfactory results. 

\section{Results and Discussion}

\begin{figure}
\centering\includegraphics[scale=0.7]{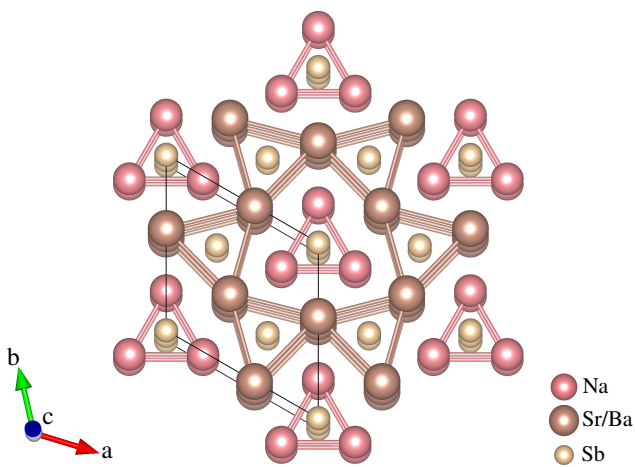}
\caption{Crystal structure of NaSrSb and NaBaSb in $P\bar62m$
symmetry, where Na atoms are located at 3\textsl{g} (\textit{x}, 0, $\frac{1}{2}$), Sr/Ba at 3\textit{f} (\textit{x}, 0, 0), whereas Sb$_1$ and Sb$_2$ at 1\textit{a} (0, 0, 0) and 2\textit{d} ($\frac{1}{3}$, $\frac{2}{3}$, $\frac{1}{2}$),
respectively.}
\label{Crystal}
\end{figure}

In this section, we will discuss our results of structural optimization,
phonon dispersions, and thermal and electrical transport.

\subsection{Crystal Structure and Phonon Dispersions}

Here, we discuss the crystal structure briefly as the
structure prototype is well explored and can be seen in
the literature \cite{Oliynyk17, Hoffmann01}. The Na\textit{X}Sb (\textit{X} = Sr, Ba)
crystallize in non-centrosymmetric hexagonal symmetry $P\bar62m$.
The traditional way of visualizing the structure is to consider
the two types of trigonal prisms, i.e., one of Na atoms and the
other of Sr/Ba stationed along the \textit{c}-axis, Fig.~\ref{Crystal}.
However, the Na prisms are well separated by Sr/Ba prisms.
The symmetry independent Sb atoms (Sb$_1$ and Sb$_2$) are
embedded at the center of the two types of prisms, respectively.
A less common alternative description of the structure
can be found in Ref. \cite{Wang23}.

\begin{table*}[]
\caption{Calculated lattice parameters, band gaps, and atomic positions
for NaSrSb and NaBaSb. The reported experimental and calculated values
from Ref.\cite{Wang23} and Ref.\cite{Chepkoech22} are mentioned in square and curly brackets, respectively.}
\centering
\begin{tabular*}{\textwidth}{l @{\extracolsep{\fill}} llllllllll}
\hline 
\hline
        &               &             &\multicolumn{2}{c}{E$_g$ (eV)}   &    \\ \cline{4-5}
System	&a (\AA)        &c (\AA)      &PBE (SOC)   &mBJ (SOC)           &\multicolumn{5}{c}{Wyckoff Positions} \\  \hline
        &               &             &            &                    &Na     &3\textsl{g}    &0.2414 &0   &1/2 \\
NaSrSb  &8.30 [8.22]    &4.89 [4.84]  &0.95 (0.82) &1.55 (1.43)         &Sr     &3\textit{f}    &0.5830 &0   &0 \\
        &\{8.30\}       &\{4.89\}     &\{0.94\}    &\{1.56\}            &Sb$_1$ &1\textit{a}    &0      &0   &0 \\
        &               &             &            &                    &Sb$_2$ &2\textit{d}    &1/3    &2/3 &1/2 \\
\hline        
        &               &             &            &                    &Na     &3\textsl{g}    &0.2410 &0   &1/2 \\
NaBaSb  &8.59 [8.48]    &5.08 [5.03]  &0.79 (0.70) &1.30 (1.20)         &Ba     &3\textit{f}    &0.5837 &0   &0 \\
        &               &             &            &                    &Sb$_1$ &1\textit{a}    &0      &0   &0 \\        &               &             &            &                    &Sb$_2$ &2\textit{d}    &1/3    &2/3 &1/2 \\
\hline
\hline
\end{tabular*}
\label{Opt}
\end{table*}

\begin{figure}
\centering\includegraphics[scale=0.45]{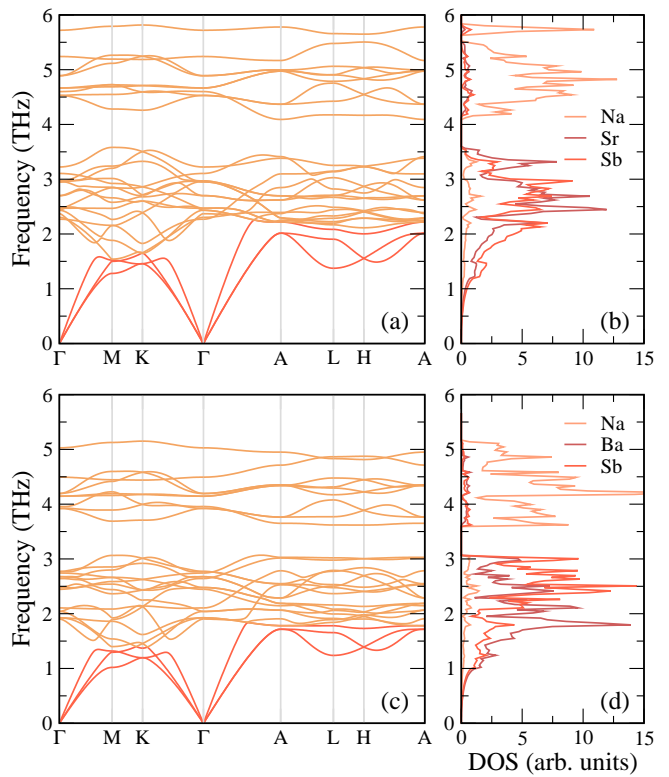}
\caption{Phonon dispersion curves [(a) and (c)]
and [(b) and (d)] partial density of states of NaSrSb and
NaBaSb, respectively.}
\label{Phonons}
\end{figure}

We optimized the crystal structure using the experimental inputs.
For optimization, we relaxed all degrees of freedom to find the
ground state. The optimized lattice parameters and atomic positions,
along with experimental and previously calculated values, are listed
in Table~\ref{Opt}. The calculated and experimental lattice constants
agree with a maximum discrepancy of less than 1.5\%.
Nonetheless, our results are consistent with previous calculations
\cite{Chepkoech22}.
Note that our values are slightly higher, which
can be attributed to the overestimation of lattice constants by GGA
\cite{Stampfl99}.
Utilizing the optimized parameters, we checked the systems'
dynamic stability through phonon calculations, as shown in
Fig.~\ref{Phonons}.

The real phonon frequencies signify the dynamic stability of
the systems. The atom-projected phonon density of states 
show that phonon modes
are categorically contributed by atoms according to their masses,
e.g., heavier atoms (Sr/Ba and Sb) predominantly contribute to
the acoustic modes and mid-frequency region phonons, whereas the
Na atoms contribute abundantly to the high frequency optical modes.
The nature of the phonons of the two systems is expectedly similar
owing to the same crystal structure and similar compositions.
The phonon band gap between low-lying and high-frequency optical
modes in both systems is worth noticing. The origin of the phonon gap
can be attributed to large difference in atomic masses (Na and Sr/Ba). An additional phonon gap is observed around 5.5 THz in NaSrSb, which is
missing in its counterpart NaBaSb. Further, the phonons span up to around
5.82 THz and 5.16 THz for NaSrSb and NaBaSb, respectively, which suggests
the flat nature of phonon bands. Since the group velocity is the
derivative of phonon frequency with respect to the wave vector,
we anticipate low group velocities in both materials. Besides, 
such low-lying phonon modes hint at low Debye temperature,
suggesting low lattice thermal conductivity in NaSrSb and
NaBaSb \cite{Li22, Chepkoech22}. Therefore, it will be interesting to see the trend
of lattice thermal conductivity, as discussed in the following
section.

\subsection{Lattice Thermal Conductivity}

\begin{figure}
\centering\includegraphics[scale=0.40]{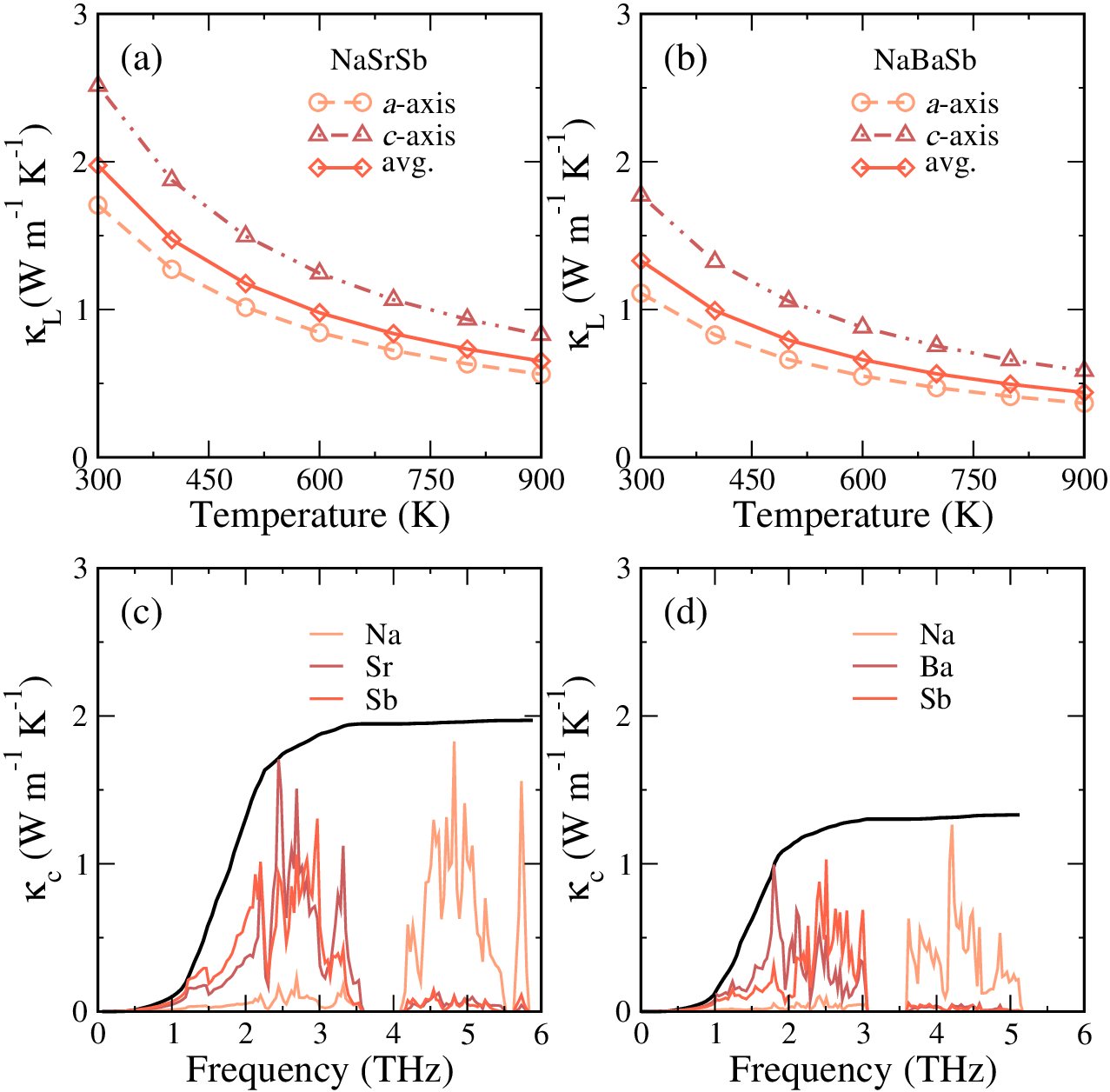}
\caption{[(a) and (b)] Lattice thermal conductivity
as a function of temperature and [(c) and (d)] cumulative lattice
thermal conductivity at 300~K as a function of phonon frequency for
NaSrSb and NaBaSb, respectively.}
\label{Kappa}
\end{figure}

We calculated the lattice thermal conductivity, $\kappa_L$, using
both the single mode relaxation time approximation (RTA) and the
direct solution of the linearized phonon Boltzmann equation (LBTE)
\cite{Chaput13}.
We found a fairly good agreement between the two methods.
The calculated $\kappa_L$ as a function of temperature for NaSrSb
and NaBaSb along the \textit{a}-axis, \textit{c}-axis, and the average
values obtained by arithmetic average along different
axes are shown in Fig.~\ref{Kappa}.
It can be seen that the $\kappa_L$ along \textit{c}-axis
overshadows the values along \textit{a}-axis. It is interesting
to observe that the room temperature $\kappa_L$ along \textit{a}-axis
are 1.7 and 1.1 W m$^{-1}$ K$^{-1}$ and along the \textit{c}-axis
are 2.5 and 1.7 W m$^{-1}$ K$^{-1}$, whereas the average $\kappa_L$
ranges 1.9--0.6 and 1.3--0.4 W m$^{-1}$ K$^{-1}$ in the temperature
region 300--900~K for NaSrSb and NaBaSb, respectively. Altogether,
the $\kappa_L$ is higher for NaSrSb than NaBaSb. Interestingly, the $\kappa_L$ values for the two materials are substantially low and are of the same order as some other low $\kappa_L$ reported Zintl phases such as Ca$_5$Al$_2$Sb$_6$ \cite{Toberer10}, Sr$_3$AlSb$_3$ \cite{Zevalkink13}, and BaCuSb \cite{Zheng22}.

To understand the fundamental reasons for such low $\kappa_L$,
we investigated different aspects of thermal properties. First,
we look into the atomic contribution towards $\kappa_L$, which is
analyzed using the calculated cumulative lattice thermal conductivity
as a function of phonon frequency. As discernible from Fig.~\ref{Kappa}(c)
and (d), the phonon modes less than $\sim$2.3 THz and $\sim$1.9 THz contribute
approximately $\sim$84\% and $\sim$79\% to the lattice thermal
conductivity of NaSrSb and NaBaSb, respectively. Now these phonon
modes are predominantly occupied by Sr and Sb in NaSrSb, whereas Ba
has the major contribution in the case of NaBaSb. Further, the low
frequency optical modes lying before the phonon band gap have some
contribution towards $\kappa_L$. The high frequency phonon modes
above the phonon gap, contributed mainly by Na, have negligible
contributions. Thus, the major contribution to
heat transport comes from Sr/Sb in NaSrSb, whereas Ba is primarily
responsible in the case of NaBaSb.

\begin{figure}
\centering\includegraphics[scale=0.40]{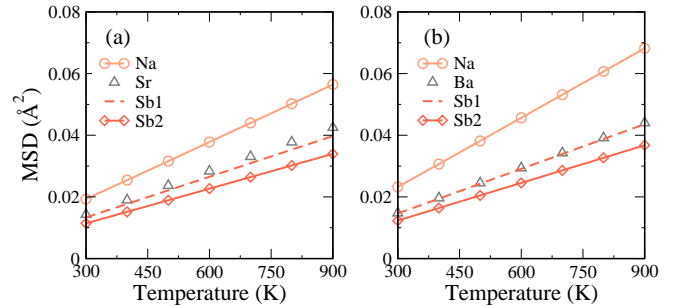}
\caption{Mean square displacements of (a) NaSrSb and
(b) NaBaSb as a function of temperature.}
\label{MSD}
\end{figure}

Now we turn to understand the role of atomic vibrations towards
$\kappa_L$ with the help of mean square displacement (MSD) of atoms.
The large MSD of atoms around the equilibrium position indicates
low $\kappa_L$ \cite{Sales99}. The Zintl phase TlInTe$_2$ is reported to have
remarkably low $\kappa_L$ on account of its large atomic displacement
parameter, i.e., 0.07 \AA$^2$ at 300~K \cite{Jana17}. We found that the MSD at 300~K
of the major contributors to $\kappa_L$, i.e., Sr/Sb in NaSrSb and Ba in
NaBaSb, is around 0.01 \AA$^2$, as can be seen in Fig.~\ref{MSD}. The
values are not significant compared to TlInTe$_2$, nevertheless, 
they still suggest reasonable vibration of atoms. Nonetheless, the two
inequivalent Sb atoms have slightly different MSD values. It is
interesting to note that the MSD of Sb$_1$ and Sr/Ba are pretty
similar throughout the temperature range, indicating bonding
between Sb$_1$-Sr/Ba. Further Na has a high MSD as compared to rest
of the atoms. This suggests the rattling behavior of Na atoms
in the cage like structure of Sb$_1$-Sr/Ba, consistent with the
Zintl phase concept as predicted in the previous findings \cite{Wang23}.

\begin{figure}
\centering\includegraphics[scale=0.40]{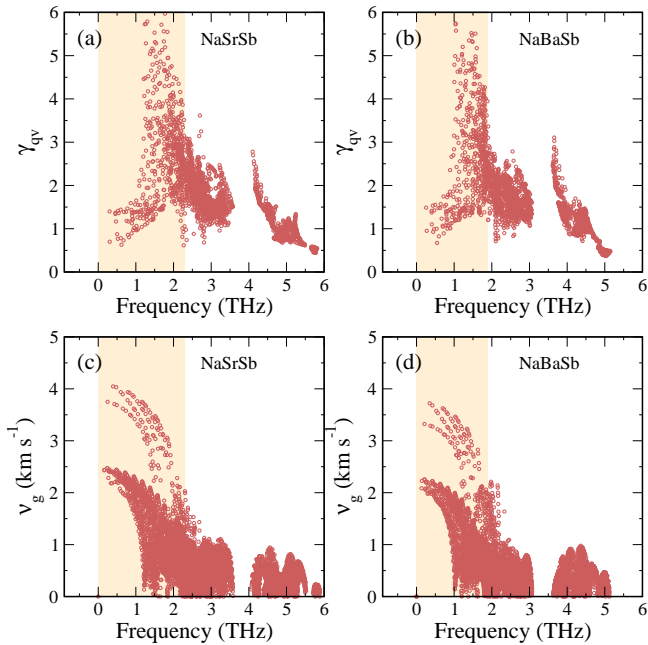}
\caption{[(a) and (b)] Mode Gr{\"u}neisen parameter
and [(c) and (d)] group velocity as a function of phonon frequency
of NaSrSb and NaBaSb, respectively. The shaded region corresponds
to phonon modes responsible for the majority of lattice thermal conductivity.}
\label{GV}
\end{figure}

The rattling of Na atoms in the Sb$_1$-Sr/Ba framework suggests
anharmonicity in the two materials, which is another descriptor
of $\kappa_L$. The anharmonicity in a crystal lattice is quantized
by the Gr{\"u}neisen parameter ($\gamma$), which measures
the deviation of phonon frequencies with respect to volume.
We observe from Fig.~\ref{GV}(a) and (b) that the majority of
phonon modes have $\gamma$ $>$ 1. The average $\gamma$
values for the phonon modes (shaded region) contributing
most to $\kappa_L$ are 1.8 and 1.5, respectively, for
NaSrSb and NaBaSb. If compared with some reputed thermoelectric
materials such as PbTe ($\gamma$ $\sim$ 1.4) \cite{Wang11} and BiCuSeO ($\gamma$ $\sim$ 1.5) \cite{ZhaoBCS}, we find that NaSrSb and NaBaSb exhibit significant lattice
anharmonicity and could be a sign of low $\kappa_L$. 

\begin{figure}
\centering\includegraphics[scale=0.40]{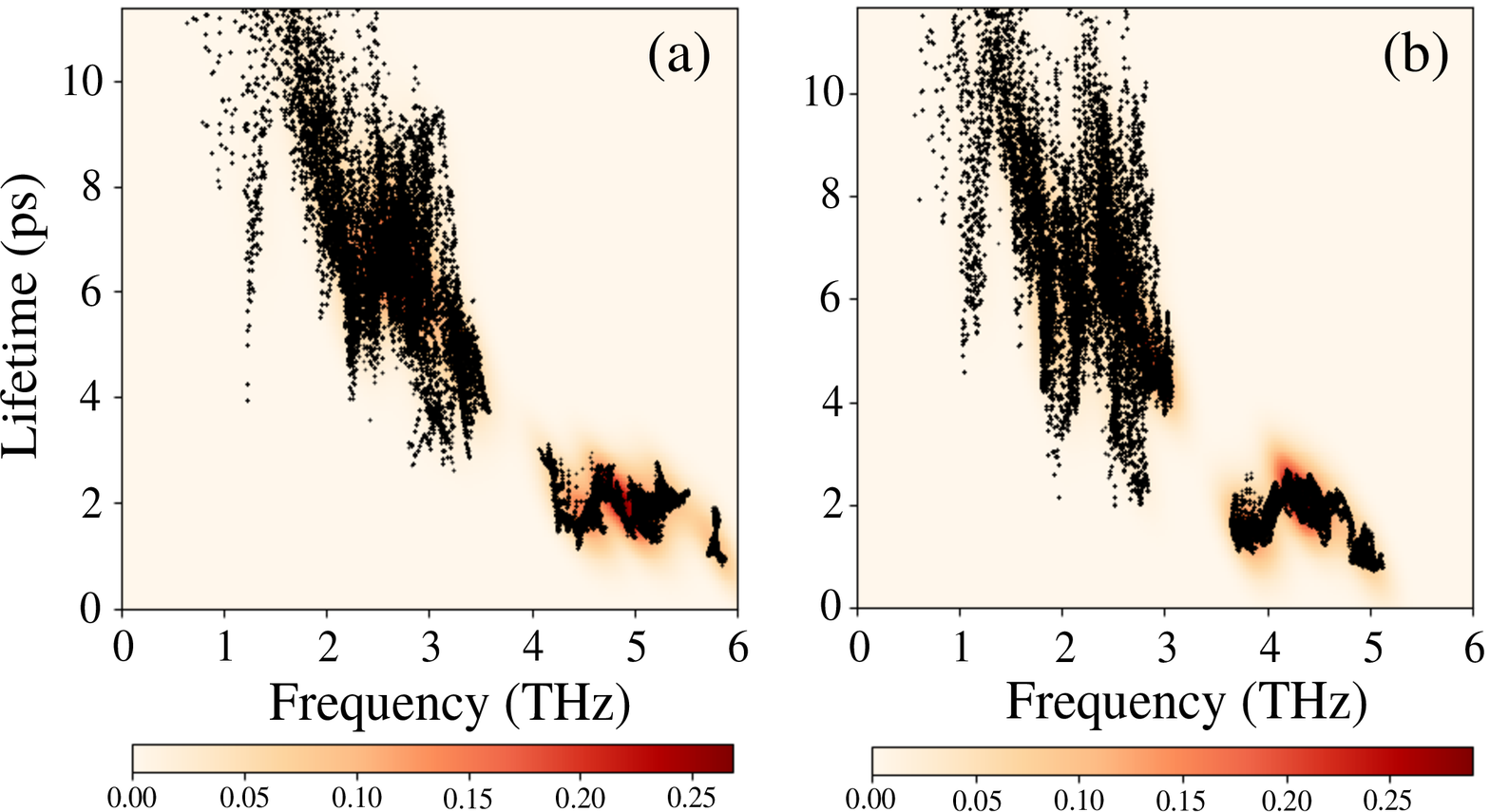}
\caption{Phonon lifetime as a function of phonon
frequency of (a) NaSrSb and (b) NaBaSb. The color bar indicates
the density of phonon modes.}
\label{Lifetime}
\end{figure}

Let us now focus on the other aspect of Fig.~\ref{GV},
i.e., phonon group velocities. As discussed before, the relatively
flat nature of phonon modes suggests low group velocities in these
materials. This is also evident from Fig.\ref{GV}(c)
and (d) which shows majority of phonon modes responsible for $\kappa_L$
exhibit group velocities less than 2.5 km s$^{-1}$. In fact,
the average group velocity of such phonons is only 0.96 and
0.84 km s$^{-1}$ for NaSrSb and NaBaSb, respectively. These 
values are strikingly low compared to BiCuSeO (2.0 km s$^{-1}$) \cite{ZhaoBCS} and SnSe (3.1 km s$^{-1}$) \cite{Guo15}.
The low group velocities of NaBaSb with respect to NaSrSb also
account for its low $\kappa_L$.
Further, the phonon lifetimes of the two materials are fairly similar
as apparent from Fig.~\ref{Lifetime}. The phonon lifetime
in both materials ranges approximately 0.5 to 11 ps, which are relatively
short compared to SnSe \cite{Guo15}.

To sum up, the short phonon lifetimes, low group velocities,
and lattice anharmonicity accounts for low $\kappa_L$ in
NaSrSb and NaBaSb. Motivated by promising lattice dynamics,
we study the electronic structure and electrical transport
properties in the following section. 

\subsection{Electronic Structure and Thermoelectric Properties}

\begin{figure}
\centering\includegraphics[scale=0.45]{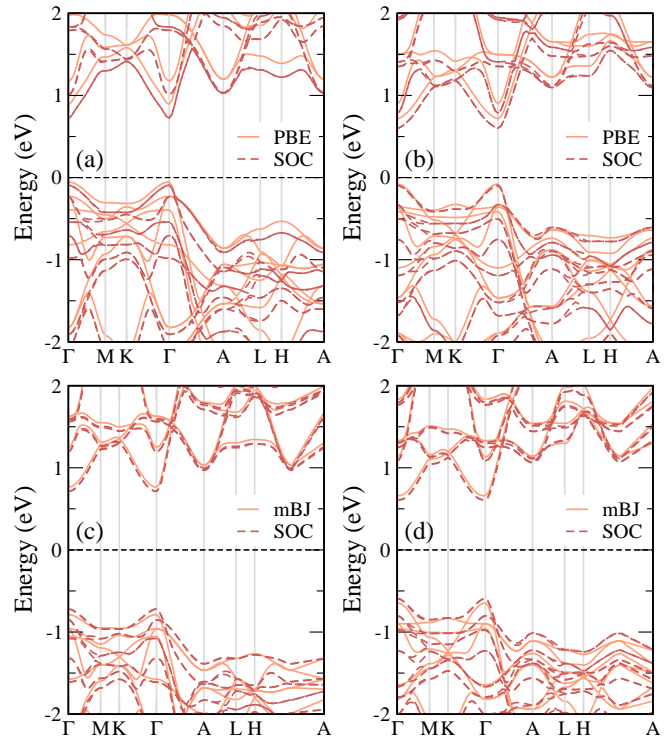}
\caption{Electronic band structure of [(a) and (c)] NaSrSb and [(b) and (d)] NaBaSb.}
\label{Bands}
\end{figure}

The calculated electronic band structure of NaSrSb and NaBaSb are
shown in Fig.~\ref{Bands}.  Both materials
are direct band gap semiconductors with valence band maximum and
conduction band mininimum stationed along $\Gamma$-point.
As the GGA-PBE is known to seriously
underestimate the band gap, we also employed mBJ potential for
calculating the electronic structure. The band gap is certainly
improved for both materials, as seen from the electronic
structure (Fig.~\ref{Bands}), and can be noted in Table~\ref{Opt}.
Further, as stated before, we considered the spin-orbit
coupling since the chosen materials comprise heavy elements. As
expected, the band gap has reduced with spin-orbit coupling due
to the shifting of bands. Importantly, the curvature and slope of the
bands near the Fermi level have not changed much, thereby
unlikely to greatly impact the transport properties. Thus, the electrical transport properties are calculated in non-relativistic
manner. However, we have exclusively used mBJ potential in order to
obtain more reliable properties. 

\begin{figure}
\centering\includegraphics[scale=0.4]{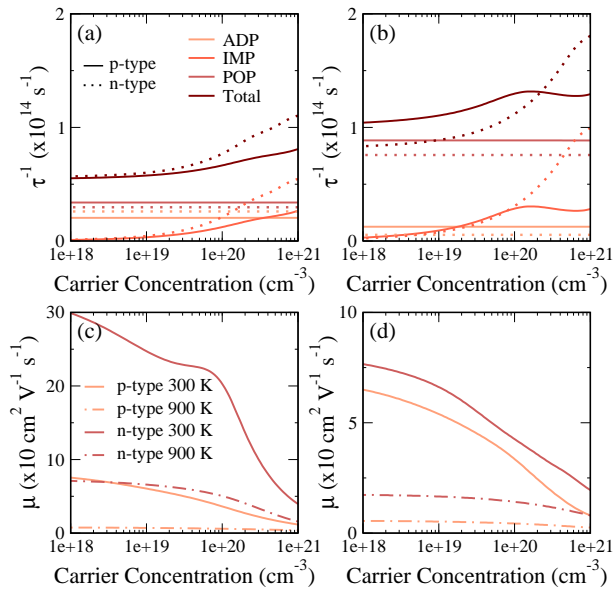}
\caption{[(a) and (b)] Scattering rates at 300~K and [(c) and (d)]
mobility at 300~K and 900~K of NaSrSb and NaBaSb, respectively, as a function of 
\textit{p}-type and \textit{n}-type carrier concentration.}
\label{Scat}
\end{figure}

We have computed the scattering rates at different temperatures and
carrier concentrations to evaluate the electrical transport properties,
considering three different scattering mechanisms, viz. ADP, IMP, and POP.
The calculated scattering rates (1/$\tau$) of charge
carriers are shown in Fig.~\ref{Scat}(a) and (b). For clarity, we have shown the results
of scattering rates only at 300~K. Note that the trend of the scattering rate
remains the same at higher temperatures, however, the magnitude is certainly on
the higher side. The ADP and POP scattering rates do not vary with the chosen
carrier concentrations, and the latter has the dominating values. Meanwhile,
IMP scattering rates gradually increases with carrier concentration and
governs the trend of the net scattering rate. Taken altogether, the 
electrons have a higher scattering rate for NaSrSb, whereas holes
have a higher scattering rate for NaBaSb until the carrier concentrations
beyond 10$^{20}$ is reached. 

Further, to understand the transport properties better, we have
shown the mobility ($\mu$) of charge carriers as a function of
carrier concentration at 300 and 900~K in Fig.~\ref{Scat}(c) and
(d). In general, the mobility of charge carriers is broadly a
consequence of the scattering rate and dispersion of bands \cite{Natarajan23}.
As discernible from the figure, the mobility of electrons is
higher than that of holes for both NaSrSb and NaBaSb at 300 and
900~K. This can be attributed to the more dispersive nature of
the bands in the conduction region. On the other hand, the mobility
at 300~K is more as compared to 900~K for either type of charge
carriers, which is essentially an effect of increased scattering
with more thermal energy. Notably, the mobility of electrons at
300~K is dramatically higher than holes for NaSrSb. Thus,
\textit{n}-type charge carriers will likely have significantly
higher electrical conductivity, as discussed ahead. 
Next, we discuss our results on thermoelectric coefficients based
on the calculated scattering rates and mobility of charge carriers.

\begin{figure}
\centering\includegraphics[scale=0.45]{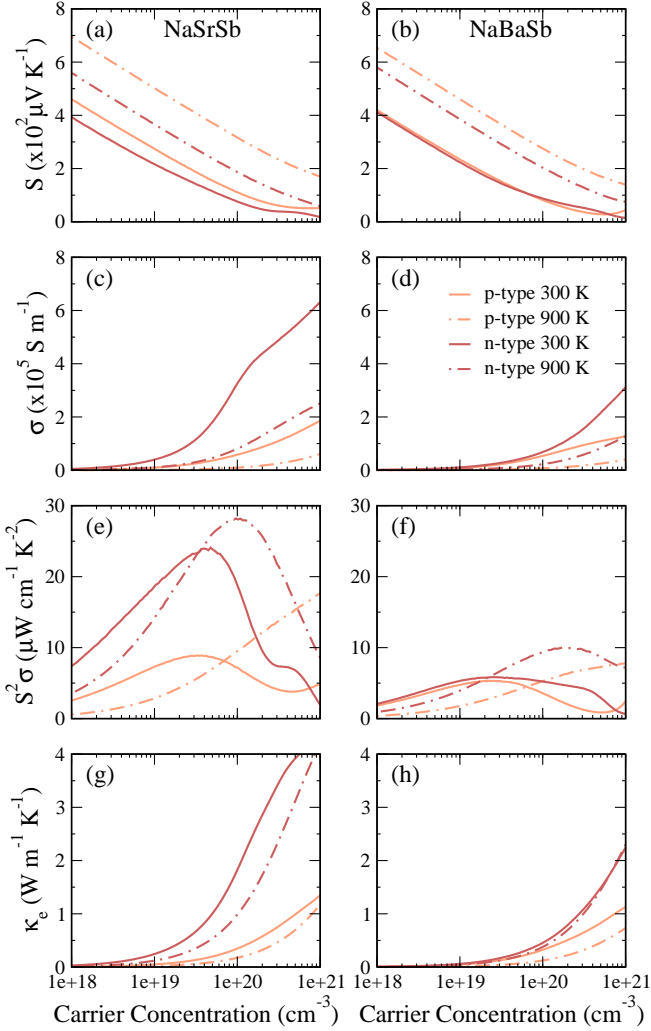}
\caption{[(a) and (b)] Seebeck coefficient, [(c) and (d)]
electrical condcutivity, [(e) and (f)] power factor, and [(g) and (h)]
electronic thermal conductivity of NaSrSb and NaBaSb, respectively,
as a function of \textit{p}-type and \textit{n}-type carrier concentration.}
\label{Transport}
\end{figure}

Figure~\ref{Transport} shows the calculated Seebeck coefficient (\textit{S}),
electrical conductivity ($\sigma$), power factor (\textit{S}$^2\sigma$), and electronic thermal conductivity ($\kappa_e$) as a function of carrier concentration at 300~K and 900~K. 
The materials under investigation are synthesized at 1173~K,
allowing us to discuss the results at room
temperature 300~K and as high as 900~K.
The directional dependent electrical transport properties
are provided in the Supplemental Material \cite{Suppl}
The computed electrical transport coefficient follows the
expected trend. The Seebeck coefficient decreases monotonously
with carrier concentration, consistent with the Mott equation
\cite{Heremans08, Cutler69}.
However, the Seebeck coefficient is dominating for \textit{p}-type
dopings. This could be attributed to flat bands near the Fermi level in the
valence band region. We found impressive values of the Seebeck coefficient
for both materials. In the case of NaSrSb, the highest obtained 
Seebeck coefficient is $\sim$460 (695) and 393 (559) $\mu$V K$^{-1}$
for \textit{p}-type and \textit{n}-type carriers, respectively, at 300~K (900~K).
The corresponding values for NaBaSb are $\sim$420 (653) and 412 (580) $\mu$V K$^{-1}$, respectively.

The electrical conductivity increases with carrier concentration because
of increased number of charge carriers, as shown in Fig.~\ref{Transport}.
The higher values of electrical conductivity can
be noticed for \textit{n}-type carrier concentrations, which may find
their origin in more dispersive bands of the conduction region. The
exceptionally high electrical conductivity of \textit{n}-type
NaSrSb at 300~K is consistent with its substantially high mobility of
charge carriers, as discussed earlier in Fig.~\ref{Scat}.
The two conflicting trends of the Seebeck coefficient and electrical
conductivity shows a peak in power factor value. We obtain higher
power factor values at 900~K for \textit{n}-type carrier concentration
in both materials, albeit, NaBaSb has almost twofold lower values.
The large value of power factor at 900~K indicates the high temperature
sustainability of these materials. For NaSrSb, a high power factor of 28.2
$\mu$W cm$^{-1}$ K$^{-2}$ at 900~K is obtained for \textit{n}-type doping
(9.6$\times$10$^{19}$ carriers cm$^{-3}$). It should be noted that the
Seebeck coefficient of the two materials is not significantly different.
The reason for the considerably high power factor of NaSrSb is its
anomalously large electrical conductivity. As far as the electronic
thermal conductivity is concerned, it mirrors the trend of electrical conductivity.

\begin{figure}
\centering\includegraphics[scale=0.43]{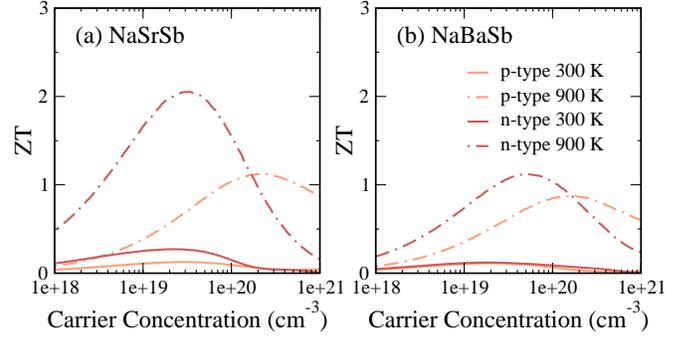}
\caption{Figure of merit at 300~K and 900~K as a function of
carrier concentration of (a) NaSrSb and (b) NaBaSb}
\label{ZT}
\end{figure}

Next, we discuss the figure of merit, which essentially describes
the performance of a thermoelectric material and is expressed as
ZT = S$^2 \sigma$T/$\kappa$ \cite{Yu18}. The results of our calculated ZT as
a function of carrier concentration at 300~K and 900~K are presented
in Fig.~\ref{ZT}. The trend of ZT is similar to the power factor due
to the conflicting parameters involved in the aforementioned expression.
The ZT values at 300~K vis-à-vis 900~K are insubstantial, therewith, we
focus on the values at the higher temperature. The ZT values are higher
for the \textit{n}-type than the counterpart \textit{p}-type carrier
concentration for both materials. We found a remarkably high 
ZT$\sim$2.0 at 900~K for an optimal carrier concentration of 3.2$\times$10$^{19}$
carriers cm$^{-3}$. The value is undoubtedly competitive with other state-of-the-art
thermoelectric materials, such as PbTe-SrTe (ZT$\sim$2.5 at 923~K)
\cite{Tan16} and SnSe (ZT$\sim$2.6 at 913~K) \cite{Zhao14}. On the other hand, the \textit{p}-type NaSrSb
exhibits a ZT$\sim$1.1 at 900~K for 2$\times$10$^{20}$ carriers cm$^{-3}$.

In the case of NaBaSb, at 900~K, we obtain a ZT$\sim$1.1 for \textit{n}-type
(5$\times$10$^{19}$ carriers cm$^{-3}$) and ZT$\sim$0.8 for \textit{p}-type
(1.8$\times$10$^{19}$ carriers cm$^{-3}$) dopings. Despite a lower $\kappa_L$
compared to NaSrSb, NaBaSb could not benefit much due to its surprisingly
low power factor, which overshadows the low $\kappa_L$. Regardless, a figure of
merit slightly higher than unity is still an appreciable number. Since the
$\kappa_L$ is reasonably low, we believe optimizing the electrical
transport parameters by means of band engineering will be a prospective
strategy in further improving the figure of merit in NaBaSb. We believe
our study would encourage experimentalists to grow these materials
for high performance thermoelectric applications.

\section{Summary}
In this paper, utilizing first-principles simulations and Boltzmann
transport theory, we have systematically investigated the ground
state properties, electronic structure, and thermoelectric properties
of two Zintl phases NaSrSb and NaBaSb. Both systems are direct band
gap semiconductors with a band gap of 0.95 and 0.79 eV, which are
improved to 1.55 and 1.30 eV, respectively, with mBJ potential.
We further checked the impact of spin-orbit coupling. The calculations
reveal the reduced band gap, however, the nature of the bands mostly
remains intact. Phonon calculations assured the dynamic stability of the systems. Interestingly, the lattice thermal
conductivity of the two systems is surprisingly low and ranges
1.9 to 0.6 and 1.4 to 0.3 W m$^{-1}$ K$^{-1}$, respectively, in the temperature region 300 to 900~K. We trace such low values to small
phonon group velocities arising from the less dispersed phonon bands,
short phonon lifetimes, and considerable lattice anharmonicity.
We further calculated the scattering rates and mobility of charge
carriers to find the electrical transport coefficients at 300 and
900~K. We have specifically emphasized transport properties at 900~K
using exclusively the electronic structure obtained by mBJ potential.
We found excellent values of the Seebeck coefficient for \textit{n}-type NaSrSb and NaBaSb, i.e., 695 and 559 $\mu$V K$^{-1}$, respectively.
Further, we found a large power factor of 28.2 $\mu$W cm$^{-1}$ K$^{-2}$
for NaSrSb, arising from its anomalously large electrical
conductivity. Taken together, we obtained a figure of merit
ZT$\sim$ 2.0 at 900~K for \textit{n}-type NaSrSb, whereas 
the figure of merit surpasses unity in the case of NaBaSb.
We believe our work could highlight the importance
of Zintl phases NaSrSb and NaBasb as potential thermoelectric
prospects.

\begin{acknowledgments}

C. K. V. and M. Z. are thankful to CSIR and SERB-DST, respectively, for their financial assistance. B. K. M. acknowledges the funding support
from the SERB, DST (ECR/2016/001454). The calculations are performed
using the High Performance Computing cluster, Padum, at the Indian
Institute of Technology Delhi.

\end{acknowledgments}

\bibliography{ref}

\end{document}